\documentstyle[11pt]{article}

    \advance\textheight by \topskip

\begin{document}

\begin{center}
{\huge \bf The method of expansion of Feynman integrals} \\[10mm] 
  S.A. Larin \\ [3mm]
 Institute for Nuclear Research of the
 Russian Academy of Sciences,   \\
 60th October Anniversary Prospect 7a,
 Moscow 117312, Russia
\end{center}

\vspace{30mm}

\begin{abstract}
The method of expansion of integrals in external parameters is
suggested. It is quite universal and works for Feynman integrals
both in Euclidean and Minkowski regions of momenta.
\end{abstract}

\newpage

During the last two decades different techniques were 
developed \cite{asex1}-\cite{asex10}
for asymptotic expansions of Feynman integrals in 
quantum field theory. These techniques allow to perform practical
calculations when exact integrations are not possible.
In the present paper we suggest
the new effective method of expansion of integrals in external
parameters -- 'the method of cancelling factors'.
It is quite universal and works for Feynman integrals 
both in Euclidean and Minkowski regions of momenta.

To demonstrate the essence
of the method of cancelling factors we begin with the following
simple integral
\begin{equation}
\label{example11}
\int_{0}^{1}\frac{dx}{(x+t)(1+x)}=\frac{\ln(1+t)-\ln(t)-\ln(2)}{1-t}.
\end{equation}
We want to expand this integral in small external parameter $t$
before the integration. The naive expansion of the integrand in the
Taylor series in $t$ does not work 
since it produces the non-integrable singularities at $x=0$. 

To get the correct expansion let us distinguish two factors in the integrand :
the expanding factor $\frac{1}{x+t}$ (the expansion of this factor
initiates the expansion of the whole integral) and the cancelling factor
$\frac{1}{1+x}$ (this factor is used to cancel singularities arising
in the expansion of the expanding factor) .
We subtract from and add to the cancelling factor its Taylor series in
$x$ up to some power $n$ 
\begin{equation}
\label{example12}
\int_{0}^{1}\frac{dx}{(x+t)(1+x)}
= \int_{0}^{1}dx\frac{1}{x+t}\left[\frac{1}{1+x}-\sum_{j=0}^{n}(-x)^j\right]+
\int_{0}^{1}dx\frac{1}{x+t}\sum_{j=0}^{n}(-x)^j.
\end{equation}

In the first
integral of the right hand sight of eq.(\ref{example12})
we can now safely perform the Taylor expansion of
 the expanding factor $\frac{1}{x+t}$ 
in $t$ up to and including the order $t^n$.
This will not generate anymore non-integrable singularities at $x=0$ since 
the factor in the square brackets 
(the subtracted cancelling factor)
has the behavior 
$O(x^{n+1})$  and thus suppresses singularities arising 
in the expansion of $\frac{1}{x+t}$.

Finally we get the desired expansion
\begin{equation}
\label{example13}
\int_{0}^{1}\frac{dx}{(x+t)(1+x)}
=\int_{0}^{1}dx\sum_{k=0}^{n}\frac{(-t)^k}{x^{k+1}}
\left[\frac{1}{1+x}-\sum_{j=0}^{n}(-x)^j\right]+
\int_{0}^{1}dx\frac{1}{x+t}\sum_{j=0}^{n}(-x)^j +
\end{equation}
\[
O(t^{n+1}\ln t).
\]
In each term of this expression some factor is expanded and 
integrations reproduce the expansion in $t$ of the exact result in the
right hand side of eq.(\ref{example11}).

Let us generalize the above considerations.
For a given integral the method of cancelling factors distinguishes 
the expanding factor (which will be expanded in external parameters
of the integral) and several (one or more) cancelling
factors (which will be used to cancel singularities in the expansion
of the expanding factor, the number of cancelling factors
is determined by the necessity to suppress all arising
singularities). For each cancelling factor 
one adds and subtracts its expansion in the integration variable
up to the necessary order at some singular point 
(the point where singularities appear in the expansion of
the expanding factor).
At last in the term containing the product of
subtracted cancelling factors
one expands the expanding factor
in external parameters up to the necessary order without
generating non-integrable singularities.

Let us now apply the method of cancelling factors to Feynman 
integrals in quantum field theory.
To regularize divergent integrals we will use dimensional
regularization \cite{dr} for convenience (but the method
is regularization independent). 
The dimension of the momentum space is defined as
$D=4-2\epsilon$ where $\epsilon$ is the parameter 
defining the deviation of the dimension from its physical 
value 4. 

We consider first the expansion at large
external momentum squared $q^2$ of the following one-loop Feynman
integral of the propagator type
\begin{equation}
\label{largeq1}
\int  \frac{d^{D}k}{(k^2-m_1^2+i0)[(k+q)^2-m_2^2+i0]},
\end{equation}
where $k$ is the integration momentum, $m_1$ and $m_2$ are the masses
of the propagators. Below in the paper we will omit the 'causal' $i0$
for brevity.

The expansion in large $q^2$ means alternatively
the expansion in small $m_1$ and $m_2$. We can not just naively expand
the integrand in Taylor series in $m_1^2$ and $m_2^2$. Such an
expansion produces incorrect result since it
generates infrared singularities at $k=0$ and $k+q=0$.
(At first glance the singularities are generated at $k^2=0$ and $(k+q)^2=0$
but it is known that Feynman integrals with one external momentum
can be always treated in Euclidean region of momenta where conditions
$k^2=0$ and $k=0$ are equivalent.)
To get the correct expansion let us distinguish two factors in the integrand :
the expanding factor $\frac{1}{k^2-m_1^2}$  and the cancelling factor
$\frac{1}{(k+q)^2-m_2^2}$.
We subtract from and add to the cancelling factor its Taylor series
in $k$:
\begin{equation}
\label{largeq2}
\int  \frac{d^{D}k}{(k^2-m_1^2)[(k+q)^2-m_2^2]} =
\int d^{D}k \frac{1}{k^2-m_1^2}\left[\frac{1}{(k+q)^2-m_2^2}
-T_k^{2n_1}\frac{1}{(k+q)^2-m_2^2}\right]+
\end{equation}
\[
\int d^{D}k\frac{1}{k^2-m_1^2}T_k^{2n_1}\frac{1}{(k+q)^2-m_2^2},
\]
where
\[
T_k^{2n_1}\frac{1}{(k+q)^2-m_2^2}=\sum_{j=0}^{2n_1}
\frac{\partial^{j}}{\partial
k^{\mu_1}...k^{\mu_j}}\frac{1}{(k+q)^2-m_2^2}\vert_{k=0}
\frac{k^{\mu_1}...k^{\mu_j}}{j!}
\]
is the Taylor expansion of the cancelling factor in $k$ up to some
order $2n_1$.

In the first
integral of the right hand sight of eq.(\ref{largeq2})
we can now perform the Taylor expansion in $m_1^2$ of
 the expanding factor $\frac{1}{k^2-m_1^2}$ up to and including
the order $(m_1^2)^{n_1+1}$.
This expansion will not generate anymore infrared singularities since 
the factor in the square brackets (the subtracted cancelling factor)
behaves as 
$O(k^{2n_1+1})$ and thus suppresses the infrared singularities at
$k=0$ arising 
in the Taylor expansion 
\[
 T_{m_1^2}^{n_1+1}\frac{1}{(k^2-m_1^2)}=\sum_{j=0}^{n_1+1} 
\frac{(m_1^2)^j}{(k^2)^{j+1}}.
\]
Thus we get
\begin{equation}
\label{largeq3}
\int  \frac{d^{D}k}{(k^2+m_1^2)[(k+q)^2-m_2^2]}=
\int d^{D}k \, T_{m_1^2}^{n_1+1}\frac{1}{k^2-m_1^2} \frac{1}{(k+q)^2-m_2^2}+
\end{equation}
\[
\int d^{D}k \frac{1}{k^2-m_1^2}T_k^{2n_1}\frac{1}{(k+q)^2-m_2^2}
+O\left((m_1^2)^{n_1+2}\right),
\]
where we took into account that the term containing both Taylor
expansions $T_{m_1^2}^{n_1+1}$ and $T_k^{2n_1}$ is zero due to the known
property of the dimensional regularization to nullify
the integrals
without external parameters (massless tadpoles).
The approximation here $O\left((m_1^2)^{n_1+2}\right)$
and approximations below in the paper are written up to logarithms.

This is already a kind of expansion but we can continue further
with the expansion of the first term in the right hand side of 
eq.(\ref{largeq3}).
For this purpose it is convenient to make in this term 
the shift of integration momentum $k\rightarrow k-q $, so we get
\begin{equation}
\label{largeq4}
\int  \frac{d^{D}k}{(k^2+m_1^2)[(k+q)^2-m_2^2]}=
\int d^{D}k \frac{1}{k^2-m_2^2} T_{m_1^2}^{n_1+1}\frac{1}{(k-q)^2-m_1^2}+
\end{equation}
\[
\int d^{D}k \frac{1}{k^2-m_1^2}T_k^{2n_1}
\frac{1}{(k+q)^2-m_2^2}
+O\left((m_1^2)^{n_1+2}\right).
\]
Then in the first term the factor $\frac{1}{k^2-m_2^2}$ is considered
as the expanding factor and the factor
$T_{m_1^2}^{n_1+1}\frac{1}{(k-q)^2-m_1^2}$ as the cancelling factor.
Again we  subtract from and add to the cancelling factor
its Taylor expansion $T_k^{2n_2}$ in $k$. Then
in the term containing the subtracted cancelling factor we can make the
Taylor expansion $T_{m_2^2}^{n_2+1}$  of the expanding factor
$\frac{1}{k^2-m_2^2}$ 
(in the same way as the 
expansion in $m_1^2$ during the derivation of eq.(\ref{largeq3})).
Finally we come to the expansion (after nullification of 
massless tadpoles)
\begin{equation}
\label{largeq5}
\int \frac{d^{D}k}{(k^2+m_1^2)[(k+q)^2-m_2^2]}=
\int d^{D}k \, T_{m_2^2}^{n_2+1}\frac{1}{k^2-m_2^2}
T_{m_1^2}^{n_1+1}\frac{1}{(k-q)^2-m_1^2}+
\end{equation}
\[
\int d^{D}k \frac{1}{k^2-m_2^2}
T_{k}^{2n_2}T_{m_1^2}^{n_1+1}\frac{1}{(k-q)^2-m_1^2}+
\]
\[
\int d^{D}k \frac{1}{(k^2-m_1^2)}T_{m_2^2}^{n_2+1}T_k^{2n_1}
\frac{1}{(k+q)^2-m_2^2}
+O\left((m_1^2)^{n_1+2}, (m_2^2)^{n_2+2}\right),
\]
where in the last term (which is nothing but the last term
in eq.(\ref{largeq4})) we applied the Taylor expansion in $m_2$
which does not effect integrations.
This result agrees with the recipe explicitly formulated in
\cite{asex7} for the large $q^2$ expansion of propagator integrals.
The new point here is the simple derivation of the expansion.

As the next application of the method
we shall consider the Sudakov formfactor \cite{sud}
which is a typically Minkowskian case not reducible to
the Euclidean space of momenta. The corresponding one-loop Feynman
integral is
\begin{equation}
\label{sud1}
\int \frac{d^Dk}{(k^2-m^2)(k^2-2p_1k)(k^2-2p_2k)},
\end{equation}
where external momenta are on mass shell: $p_1^2=p_2^2=0$.
The integral will be expanded in small mass $m^2$
which means the expansion in terms of the ratio $\frac{m^2}{q^2}$ 
where $q=p1-p2$. This expansion was obtained in \cite{asex9}.
Here we give the simple derivation of the expansion
with the method of cancelling factor.

The expansion of the integrand in $m^2$
generates infrared singularities at $k^2=0$. We distinguish here the
expanding factor $\frac{1}{k^2-m^2}$ and two cancelling factors
$\frac{1}{k^2-2p_1k}$ and $\frac{1}{k^2-2p_2k}$. For each cancelling
factor we subtract and add its expansion in $k^2$ 
\[
T_{k^2}^n
\frac{1}{k^2-2p_ik}=-\sum_{j=0}^n\frac{(k^2)^j}{(2p_ik)^{j+1}}, \, i=1,2.
\]
In this way we get
\begin{equation}
\label{sud2}  
\int \frac{d^Dk}{(k^2-m^2)(k^2-2p_1k)(k^2-2p_2k)}=
\end{equation}
\[
 \int d^Dk\frac{1}{k^2-m^2} 
\left[\left(1-T_{k^2}^n+T_{k^2}^n\right)\frac{1}{k^2-2p_1k}\right] 
\left[\left(1-T_{k^2}^n+T_{k^2}^n\right)\frac{1}{k^2-2p_2k}\right]= 
\]
\[ 
\int d^Dk\frac{1}{k^2-m^2} 
\left[(1-T_{k^2}^n)\frac{1}{k^2-2p_1k}\right] 
\left[(1-T_{k^2}^n)\frac{1}{k^2-2p_2k}\right]+ 
\]
\[
\int d^Dk\frac{1}{k^2-m^2}  
\frac{1}{k^2-2p_2k} T_{k^2}^n\frac{1}{k^2-2p_1k}
+\int d^Dk\frac{1}{k^2-m^2} 
\frac{1}{k^2-2p_1k} 
T_{k^2}^n\frac{1}{k^2-2p_2k} ,
\] 
where in the last equation
we took into account that the terms containing two factors with 
$T_{k^2}^n$ are zero. (Here is a technical subtlety. 
Dimensional regularization does not regularize individual terms in the
last equation although it regularizes the original integral
(\ref{sud1}). Strictly speaking, we
should  introduce analytic
regularization $\frac{1}{k^2-2p_ik} \rightarrow 
\frac{1}{(k^2-2p_ik)^{1+\lambda_i}},\, i=1,2$
in eq.(\ref{sud1}) in addition to
dimensional regularization, where $\lambda_i$ are the arbitrary
parameters of analytic regularization. But this technical subtlety does not
change the derivation of the expansion and the final result.)

Then in the first term of the last equation
we can expand the expanding factor as
\[
T_{m^2}^n \frac{1}{k^2-m^2} =\sum_{j=0}^n\frac{(m^2)^j}{(k^2)^{j+1}}
\]
without generating infrared singularities at $k^2=0$. This is because
the first square bracket (the first subtracted cancelling factor) behaves as 
$O\left( \frac{(k^2)^{n+1}}{(2p_1k)^{n+2}}\right)$
and the second square bracket (the second subtracted cancelling 
factor) behaves as
$O\left( \frac{(k^2)^{n+1}}{(2p_2k)^{n+2}}\right)$
at small $k^2$.
The scalar product $2p_1k$ can be small simultaneously with $k^2$
and then the first square bracket does not suppress infrared
singularities at small $k^2$.
In this case the second square bracket ensures the suppression of infrared
singularities at $k^2=0$ and vice versa.
(The scalar products $2p_1k$  and $2p_2k$
are not simultaneously small at small $k^2$ 
since the momenta $p_1$ and $p_2$ are different).
That is why we need two cancelling factors here.
Finally we get the following expansion 
(after taking into account that terms containing
two factors with Taylor expansions are zero)
\begin{equation}
\label{sud3}  
\int \frac{d^Dk}{(k^2-m^2)(k^2-2p_1k)(k^2-2p_2k)}
= \int d^Dk \, T_{m^2}^n\frac{1}{k^2-m^2} 
\frac{1}{k^2-2p_1k} 
\frac{1}{k^2-2p_2k}
\end{equation}
\[ 
+\int d^Dk\frac{1}{k^2-m^2}  
\frac{1}{k^2-2p_2k} T_{k^2}^n\frac{1}{k^2-2p_1k}
\]
\[
+\int d^Dk\frac{1}{k^2-m^2} 
\frac{1}{k^2-2p_1k} 
T_{k^2}^n\frac{1}{k^2-2p_2k} +O\left((m^2)^{n+1}\right).
\]
Here the second and third integrals are not individually regularized
by dimensional regularization as was already mentioned above 
but their sum is regularized.  

To conclude, in the present paper we described the method of
cancelling
factors for expansion of integrals in external parameters, giving
three examples of its applications.

The author gratefully acknowledges the support 
by the Russian Fund for
Basic Research under contract 97-02-17065 and by Volkswagen Foundation
under contract No. I/73611.

\end{document}